\documentclass{iaus}
\usepackage{graphicx}
\usepackage{multicol}

\title[Chemically Consistent Galaxy Models] 
{Chemically Consistent Evolutionary Synthesis Modelling of Galaxies}

\author[Fritze \& Tepper Garc\'\i a]   
{Uta Fritze$^1$ \& Thorsten Tepper Garc\'\i a$^2$}

\affiliation{$^1$University of Hertfordshire, UK, {\sl e-mail: 
ufritze@star.herts.ac.uk}\\
$^2$Universit\"at G\"ottingen, Germany, 
{\sl e-mail: tepper@astro.physik.uni-goettingen.de}}

\pubyear{2006}
\volume{235}  
\pagerange{}
\date{?? and in revised form ??}
\jname{Proceedings Galaxy Evolution Across The Hubble Time}
\editors{A.C. Editor, B.D. Editor \& C.E. Editor, eds.}

\newcommand{\ea}{\textit{et al.}}

\begin{document}

\maketitle

\begin{abstract}
We present our chemically consistent GALEV Evolutionary Synthesis models 
for galaxies and point out differences to previous generations of models 
and their effects on the interpretation of local and high-redshift galaxy data.
\end{abstract}

GALEV evolutionary synthesis models start from a gas cloud of primordial
abundances and form stars according to a given Star Formation History (SFH) and
a stellar Initial Mass Function (IMF). SFH and IMF are the only free parameters.
Models use a data base of stellar input physics to calculate both the chemical
evolution of the gas and the spectral evolution of the stellar component from
the onset of SF until the present time.
Coupled to a cosmological model, as specified by ${\rm H_0, \Omega_m,
\Omega_{\lambda}}$, GALEV models also describe the redshift evolution of 
chemical abundances and spectral properties.  
The SFHs are the basic parameter, our default IMF is Salpeter, other IMFs are
investigated for comparison. Simple Stellar Populations (SSPs) like star
clusters are formed without chemical enrichment in a single short burst ($\sim
10^5$ yr), all of their stars have the same metallicity and age. Galaxies are 
composite stellar populations with SFHs extended in time. For undisturbed galaxy
types, simplified Star Formation Rates (SFRs) are used for the respective
spectral types: SFR(t)${\rm \sim e^{-t/\tau_{*}}}$ with $\tau_{*} =
1~\mathrm{Gyr}$ for
classical Es, and SFRs coupled to the evolving gas content G, SFR(t)${\rm \sim
G(t)}$, with  characteristic timescales increasing from S0s to Sc's and const.
SFRs for Sd's, similar to the Bruzual \& Charlot, Pegase, and Starburst99
models. These SFHs in principle go back to Sandage (86) and are constrained in
detail be agreement between models and local galaxy samples/templates.
Interacting galaxies and hierarchical assembly are described via the starbursts
they trigger in our simplified 1-zone models without dynamics.

Input physics for the stars and the gas includes stellar evolutionary
tracks/isochrones (Padova vs. Geneva), stellar model atmosphere spectra and Lick
absorption features, gaseous emission in terms of lines and continuum, and
stellar yields (PNe, SNII, SNIa) for individual elements (He, C, N, O, . . .,
Fe). Complete sets of input physics are available for 5 different metallicities
[Fe/H]$=-1.7~\dots~+0.4$ for solar scaled abundances. GALEV model output is the time
evolution of Color -- Magnitude -- Diagrams (CMDs) and the time and redshift
evolution of integrated quantities: spectra (${\rm 90 \AA ~\dots~ 160 \mu m}$),
emission line strengths, luminosities and colors in various filter systems
(Johnson, HST, Washington, Str\"omgren, \dots), Lick absorption features
(Mg$_2$, Mgb, Fe5270, Fe5335, TiO1, TiO2, \dots), galaxy (or star cluster)
masses: gas and stars, M/L, ISM abundances for the above elements calculated
from a modified version of Tinsley's equations including SNIa contribution \`a
la Matteucci (carbon deflagration white dwarf binaries). {\bf All the pieces of
input physics depend significantly on metallicity and so does, of course, the
output.} This is best seen on SSPs for different metallicities: low metallicity
stellar populations are hotter, brighter, bluer, have stronger UV and ionising
fluxes, hence stronger emission lines, than high metallicity populations (Schulz
\ea{} 02). Emission line ratios of heavy element lines behave in a complex way as
a function of metallicity, therefore we use empirical line ratios (cf. Anders \&
Fritze 03). Together with stellar lifetimes, stellar yields for individual
elements also  show a complex metallicity dependence with stellar yield ratios
significantly deviating from solar ratios at low metallicities, e.g. [Mg/Fe]
increases with decreasing metallicity. Different SFHs -- short and burst-like
vs. mild and $\sim$  constant -- lead to different abundance ratios in the gas
between elements with  different nucleosynthetic origin, as e.g. [C/O], [N/O],
or [$\alpha$/Fe].  C and N originate in intermediate mass stars, Fe has
important SNIa contributions, both leading to a delayed production with respect
to the SNII products O, Mg, etc. A threshold in metallicity, below which SNIa
may be inhibited (Kobayashi \ea{} 98), would further enhance this effect. {\bf Via
the SFH, galaxy evolution and stellar evolution become intimately coupled}. In
principle, stellar evolutionary tracks, yields and model atmospheres are
required not only for various metallicities (and He contents), but also for
various abundance ratios. However, no complete grids of stellar input physics
are available yet for varying abundance ratios [${\rm [C/O],~[N/O],~[\alpha/Fe],
\dots}$]. {\bf Our chemically consistent ($=$ {\bf cc}) GALEV models follow the 
evolution of ISM abundances together with the spectrophotometric properties 
of galaxies and account for the increasing initial metallicity of successive 
generations of stars} by using for each stellar generation the set of input
physics appropriate for its initial metallicity. Models thus appropriately
account for the observed broad stellar metallicity distributions (typically
$\leq 2$ dex) in galaxies  (e.g. Rocha-Pinto \& Maciel 98, Sarajedini \&
Jablonka 05, Harris \& Harris 00) as well as for the increasing importance of
low metallicity (sub-)populations in local late-type and dwarf galaxies
(Kobulnicky \& Zaritsky 99) and in galaxies at higher redshift, in particular
the intrinsically fainter ones less accessible to spectroscopy but detected in
huge numbers in deep multi-band imaging surveys (Mehlert \ea{} 02, Pettini \ea{} 02,
Tremonti \ea{} 04).

Models clearly show that light contributions from various age and metallicity
subpopulations are very different at different wavelengths, whence the enormous
analytical power of multi-band deep photometry over a long wavelength basis and
whence significantly different metallicities are expected to be seen at different
wavelengths. In spirals the youngest stars with the highest metallicities
clearly dominate at the shortest wavelength with the result that the stellar
metallicity seen in U is higher than in V and much higher than in K where older
and less metal rich stars dominate the light. In ellipticals, in turn, the
metallicity of stars dominating in K is about 1.8 and 2 times higher than that
of stars dominating the light in V and U (cf. M\"oller \ea{} 97, Bicker \ea{} 04). 
This fact has bearing on any attempts to calibrate new stellar metallicity
indicators, e.g. in the NIR. They will inevitably show different stellar
metallicities than their optical counterparts because of the composite nature of
stellar populations in galaxies and the fact that different stellar age and
metallicity subpopulations dominate the light at different wavelengths. Models
also show that even classical initial collapse elliptical galaxies feature broad
stellar metallicity distributions, in good agreement with observations of the
stellar metallicity distributions in the halo of NGC 5128 (Harris \& Harris 00).
Therefore, elliptical galaxies cannot be adequately described/analysed by simple
(=single metallicity, single age) stellar population models. Models also show
how the stellar metallicity distributions and the light contributions of stellar
subpopulations of various metallicities have evolved with time. E.g., at a
redshift ${\rm z \sim 1}$, when galaxies have about half their present age, no
solar metallicity stars are yet formed in average luminosity Sb-type galaxies,
the bulk of their stars have $1/4$ to $1/5$ of solar metallicity (see Bicker \ea{}
04 for details). Were these galaxies at z$\ge 1$ analysed with solar metallicity
models, their SFRs, metallicities, photometric masses etc. would be seriously in
error, as are already quantities derived for local late-type and dwarf galaxies
unless their subsolar metallicities are appropriately taken into account. In
Bicker \& Fritze (05), we showed that SFRs derived from H$_{\alpha}$ and [OII]
for nearby low metallicity galaxies like IZw18 or SBS 0335 are overestimated by
factors 2 and 3, respectively, when using the standard calibrations from
Kennicutt (98) or Gallagher \ea{} (89) that are valid for solar or near-solar
metallicities. Accurate SFRs can only be derived together with metallicities
from either spectroscopy or multi-band imaging.

To obtain from the time evolution of gas abundances their redshift evolution only
requires a transformation between age and redshift, directly given by the
cosmological parameters and an assumed redshift at which SF started. Our
comparison of the chemical evolution of the ISM in spiral galaxies from cc GALEV
models with observed element abundances from Keck HIRES spectroscopy in the
neutral gas of Damped Ly$\alpha$ Absorbers (DLAs) in Lindner \ea{} (99) has shown
{\bf a)} that DLAs observed in ample numbers to redshifts z$=4$ and higher can
well be the progenitors of local spiral galaxy types Sa \dots Sd with model
abundances naturally connecting the low DLA abundances (typically 1/100 to few
tenths of solar at redshifts z$=2 - 4$) to HII region abundances in local
spirals, {\bf b)} that the DLA phase is a normal transition phase in the life of
a spiral in the sense that early-type spirals reaching high metallicities at the
same time have consumed their gas to an extent where they drop out of the high
column density DLA samples, {\bf c)} that proto-spirals at z$=2 - 4$ already
must have had 50 - 100 \% of their present-day total mass, albeit largely in the
form of gas, in agreement with dynamical DLA mass estimates from rotation
velocities by Prochaska \& Wolfe (97), Wolfe \ea{} (05), and {\bf d)} that DLA
galaxies have faint luminosities and low SFRs in agreement with recent VLT
(non-)detections. The most important conclusion from this comparison between DLA
abundances and our spiral models is that it confirms the average SFHs of our
spiral models over a lookback time to z$>4$ of more than 90\% of the Hubble time.

To transform the time evolution of spectral quantities into their respective
redshift evolution requires an evolutionary correction (accounting for the fact
that distant galaxies are seen in younger evolutionary stages), a cosmological
correction (accounting for redshifting and dimming of the galaxy light on its
way through the expanding universe) and including the effect of attenuation of
the galaxy light at wavelengths below Ly$\alpha$ due to the cumulative effect of
intergalactic neutral hydrogen HI (cf. Madau 95, Bershady \ea{} 99, Tepper
Garc\'{\i}a 06). With all these effects properly taken into account, our cc
GALEV models give a fair description of the redshift evolution of galaxies in
the Hubble Deep Field North with photometric redshifts from Sawicky \& Yee (97).
They thus allow to identify progenitors at high redshift of the different local
galaxy types and to study their respective mass assembly and chemical enrichment
histories. The agreement between spectrophotometric observations and models out
to redshifts z$\sim 4$ also confirms our model SFHs (Bicker \ea{} 04). A number of
galaxies at z$\ge 1$ bluer than our bluest Sd model and some galaxies at ${\rm
0.5 \leq z \leq 2.5}$ redder than our reddest undisturbed E model are readily
explained in terms of starburst and post-starburst models, respectively, as
shown in Fritze \& Bicker (06a). We found starbursts to be frequent at z$>1$ and
very strong at high redshifts, increasing the stellar mass by $\geq 30$ \%.

In Bicker \& Fritze ({\sl in prep.}) we calculated a large grid of cc GALEV
models for various galaxy types (E, \dots, Sd, starbursts, and post-starbursts)
and all redshifts and developed an analysis tool that, on the basis of a
$\chi^2$ method, compares the observed Spectral Energy Distribution (SED) of a
galaxy to the model grid to find {\bf not only the best but all acceptable
fits}, and determine galaxy type and photometric redshift, SFR and age, gaseous
and stellar masses and metallicities, all including their respective 1-$\sigma$
uncertainties. Comparison with the subsample of HDF-N galaxies with
spectroscopic redshifts shows that our photometric redshifts agree with the
spectroscopic ones to $\leq 5$\% (Fritze \& Bicker 06b). Doing this kind of
analysis on the basis of cc GALEV model spectra as opposed to using a locally
observed set of galaxy spectra has the advantage to consistently account for
evolutionary and cosmological effects including attenuation and for the
increasing importance of subsolar metallicity stars in distant younger galaxies.
Comparison with the same analysis done on a grid of GALEV models {\em using
solar metallicity input physics only} shows that in this case acceptable
solutions are also found, but often with wrong (type, redshift) combinations and
that ages are underestimated by typically a factor of 2, photometric masses tend
to be overestimated by factors up to 5 and more, and SFRs overestimated by
factors 2 and higher (Tepper Garc\'{\i}a \& Fritze b {\sl in prep.}).

So far, the chemical and spectral aspects of galaxy evolution have been coupled
consistently in our GALEV models, next steps will be the consistent inclusion of
dust with the dust content coupled to the evolving gas content and metallicity
and accounting for the relative distributions of dust and stars in different
galaxy types (see M\"oller \ea{} 01a, b, c for an encouraging first attempt) and
the coupling of cc GALEV models with dynamical galaxy models including gas,
stars, SF and feedback on the relevant (pc to kpc) scales. The coupling of GALEV
models into a cosmological structure formation simulation was first attempted by
Contardo \ea{} (98), albeit by that time with still insufficient resolution, in 
particular for SF and feedback. 

\vspace{-0.5cm}

\end{document}